\documentclass[12pt]{article}
\usepackage{graphicx}
\usepackage{epsf}

\newcommand{\be}[1]{\begin{equation}\label{#1}}
\newcommand{\ee}{\end{equation}}
\newcommand{\bea}[1]{\begin{eqnarray}\label{#1}}
\newcommand{\eea}{\end{eqnarray}}
\newcommand{\micron}{\,\mu m}

\begin{document}
\begin{center}
{\LARGE \bf  The relation of morphology and affinity maturation
in germinal centers}\\
\vspace{5mm}
Michael Meyer-Hermann\\
\vspace{2mm}
Institut f\"ur Theoretische Physik, TU Dresden,
D-01062 Dresden, Germany\\
E-Mail: meyer-hermann@physik.tu-dresden.de
\end{center}
\vspace{2mm}
\noindent{\bf Abstract:}
The specific morphology of germinal centers is analyzed
in the context of the optimization of the humoral immune
response.
The relevance of dark and light zones for the affinity
maturation process is investigated in the framework
of a theoretical model for the germinal center reaction. 
Especially,
it is shown that an intermediate appearance of dark zones
in germinal center reactions is advantageous for the process
of antibody optimization. Methodological aspects are
discussed.
%


\section{Introduction}

An important part of the humoral immune response is the
germinal center (GC) reaction. GCs are responsible for
an optimization process of antibodies with respect to
a specific antigen. This process is called affinity
maturation: During the GC reaction new plasma cells 
are generated which secrete antibodies of considerably
higher affinity to the antigen.

The GC reaction is initiated by antigen-activated
B-cells that migrate into the follicle system. Here,
they start to proliferate in the environment of
follicular dendritic cells (FDCs). The initiation
is believed to be of oligoclonal character, i.e.~the
number of seeder B-cells is small and of the order of
three \cite{kro87}.
After three days of fast monoclonal expansion
-- the total number of proliferating B-cells (centroblasts)
reaches about $12000$ -- a phase of somatic
hypermutation is started \cite{jac93}.
The diversity of encoded antibodies is enhanced in this way.
The centroblasts differentiate into antibody-presenting 
centrocytes \cite{han97} and an apoptotic process is
initiated.
However, they have the possibility to get into interaction
with the antigen-presenting FDCs and with T-helper cells.
It is believed that this interaction depends on the
affinity of antibody and antigen, and that these centrocytes
which successfully bind the antigen are rescued from
apoptosis \cite{liu89,eij01}.
This provides a more-step selection process \cite{lin97}
of these B-cells with high affinity to the antigen.
Positively selected B-cells further differentiate into
plasma- and memory-cells (shortly denoted as output cells). 
In this way the answer of
the immune system is optimized with respect to the
antigen.

GCs develop a very specific morphology \cite{liu91,cam98}. 
The proliferating
and mutating centroblasts are collected in the dark zone while
centrocytes and FDCs build up the light zone. 
It is unclear how long the dark zone remains present
during a GC reaction. The total duration of a GC reaction
is about $21$ days \cite{liu91}.
Dark zones have been observed
to appear at day $4$ and to vanish at day $8$ \cite{cam98}.
However, there also exists evidence for dark zones of
longer duration \cite{liu91}.

In the present article a possible correlation of the GC
morphology and affinity maturation is investigated.
in the framework of a mathematical model \cite{mey02} 
which considers the spatial distribution of cells
in the GC for the first time.
The assumptions made in this model will be compared
to other model architecture.
The postulates of the model are summarized and 
shortly described in Sec.~\ref{model}. Note, that
all parameters of the model are quantitatively
determined in narrow connection to experiment.
An analysis of the robustness of the results has been
performed previously \cite{mey02,mey01}. Here, we aim to
report one major outcome of the model (Sec.~\ref{result}). 
An interpretation
and possible implications of the results 
as well as methodological aspects
are discussed in Sec.~\ref{discuss}.


\section{The scheme of the model}
\label{model}

A short description of a previously introduced
mathematical model for the morphological organisation
and cell dynamics of the GC \cite{mey02} is provided in this
section.
The GC is simulated on an equidistant two-dimensional
lattice with lattice constant $10\micron$. This corresponds
to the average cell diameter of B-cells in GCs.
The radius of the lattice is $220\micron$, corresponding
to a typical radius of a GC.
Each lattice point can be occupied by exactly
one centroblast, centrocyte, or output cell. 
All cells actively and isotropically move on the
lattice. 
The diffusion constants are
adapted corresponding to the different diameters of
centroblasts and centrocytes \cite{liu94}.
FDCs are represented by a soma at one lattice point and
four (in 2 dimensions) dendritic arms of $30\micron$ length.
FDCs are assumed to be immobile reflecting a rather stable
FDC network observed in experiment.

It has been previously shown that the development of dark zones
can be explained on the basis of non-local 
cell-cell interactions \cite{mey02}.
Such an interaction
may be provided by a chemotaxis gradient which acts on the
motility of centrocytes and stems from
FDC and/or T-cells and/or naive B-cells in the mantle 
zone \cite{bey02}.
Another possibility (which will be used in this work)
is a diffusing signal molecule which
is produced by FDCs and bound by centroblasts \cite{mey02}. 
Note, that this implies a separation of signals acting on
proliferation and differentiation of centroblasts, as has
been proposed in corresponding experiments \cite{han95b}.
The signal molecules are clustered in quanta that diffuse
on the lattice according to a classical diffusion equation.
The diffusion is not influenced by the presence of cells at
the same lattice point. One quantum corresponds to the
signal concentration that is necessary to initiate
the centroblast differentiation process 
into centrocytes. 
Using this non-local concept an intermediate dark zone
is produced \cite{mey02}. The duration of the dark zone basically is
dependent on the amount of secreted signal molecules and
its diffusion constant. The ratio of centroblast
differentiation and proliferation rates
changes the duration of the dark zone as well.
However, this ratio also has influence on the total life time
of the GC as a whole and, therefore, is determined independently.

The affinity of the encoded antibodies to the antigen
is formulated with the well known shape space concept \cite{per79}.
Each type of antibody is represented on a 
four-dimensional lattice
which is ordered in such a way, that neighboring points
have similar affinity to the antigen. A hypermutation is
represented by a jump to a neighbor point.
The affinity between the antibodies on a centrocyte and
the antigen on an FDC is modeled by a gaussian affinity
weight function centered at the optimal antibody type \cite{mey01}.

The dynamical properties of the different cell types
on the lattice are summarized in the following.
The parameter values have
been quantitatively determined using experimental constraints.
In many cases, indeed, the parameters were directly accessible
in experiments. Others had to be determined indirectly 
using experimental observations of the general GC
properties (for more details we refer to \cite{mey02,mey01}):\\[1ex]
{\bf Centroblasts} proliferate with a (constant) rate of 
$1/6~hr$ \cite{han64}.
At each division a somatic hypermutation occurs with
the probability $1/2$ \cite{nos91}.
They differentiate in dependence
on a differentiation signal that is secreted by the FDCs,
and diffuses over the lattice.
The differentiation process is activated when a centroblast
meets a threshold quantum of differentiation signal at the
same lattice point.
Activated centroblasts differentiate
with a rate of $1/6~hr$ into centrocytes \cite{cho00}.
A finite life time of centroblasts is not imposed. However,
the effective life time is shorter than $1$ day due to centroblast
differentiation.\\[1ex]
{\bf Centrocytes} die with a rate of $1/7~hr$ \cite{liu94}.
They bind to the FDCs according to
the affinity to the antigen according to an affinity
weight function. Bound centrocytes
remain bound for $2~hr$ \cite{eij99}.
They are thought to
be rescued from apoptosis during this time.
Positively selected centrocytes further differentiate
with a rate of $1/7~hr$
into either re-proliferating centroblasts (with probability
$80\%$) \cite{mey01}, 
or into output cells (with probability $20\%$). 
The differentiation into
output cells is delayed by $48~hr$ with respect to
the starting time of hypermutation \cite{mey01},
i.e.~it starts at day 6 of the GC reaction.\\[1ex]
{\bf Output cells} leave the GC by diffusion and do not
further interact with other cells in the GC.\\[1ex]
{\bf Dead cells} are eliminated from the GC.

The simulations are started with $3$ randomly distributed seeder
B-cells and $20$ FDCs.
The fact that centroblasts proliferate at least
in parts outside the FDC network during their proliferation
phase turned out to be a necessary requirement for the
development of dark zones \cite{mey02}.
This is ensured by a random distribution of the FDCs
on $70\%$ of the (maximum) GC volume.
The seeder cells are of low but non-vanishing
affinity to the antigen. They can reach the optimal antibody-type
with $5$ to $10$ mutations \cite{wed97}.
The simulations are insensitive to a change of the
time-step-width which is $0.004~hr$ for the presented results.
In a stochastic model the outcome of the
simulation depends on the used generator of random numbers
and on its initialization.
Therefore, the results are given with a standard deviation corresponding
to this uncertainty.


\section{Optimisation of affinity maturation}
\label{result}

At first some basic properties of the GC reaction are reported
as generated by the model simulation. 
Assuming the already introduced centroblast differentiation signal
(secreted by FDCs and bound by centroblasts)
a dark zone develops. It appears at day $4$ of
the reaction, and remains stable for a duration
that depends on the production rate of the signal molecules.
In order to simulate different durations of dark zones, the
signal production rate is varied. For each production rate
the differentiation rate of centroblast is adjusted correspondingly
(within physiological constraints)
in order to ensure a comparable final state of the GC 
after 21 days of the reaction.
The development of the dark zone as well as its depletion are
not principally affected by the variation of other parameters
within their physiological constraints. For example a smaller
proliferation rate basically scales the whole GC reaction
without changing the general behavior.

The time course of the total GC volume is in accordance with
experimental observations \cite{liu91,hol92}
provided that the dark zone
vanishes between day $6.5$ and $10$ of the GC reaction
(i.e.~that the signal production rate is chosen correspondingly):
After an exponential increase of the total cell population, a
maximum is reached after $4$ days of the reaction.
The total cell population then is diminished steadily
until the end of the reaction after $21$ days. 
At this time only about $50$ proliferating B-cells remain 
in the GC \cite{liu91}.
Taking these results together, the general GC morphology 
is well described by the model results.

\begin{figure}[ht]
\begin{center}
\includegraphics[height=9cm]{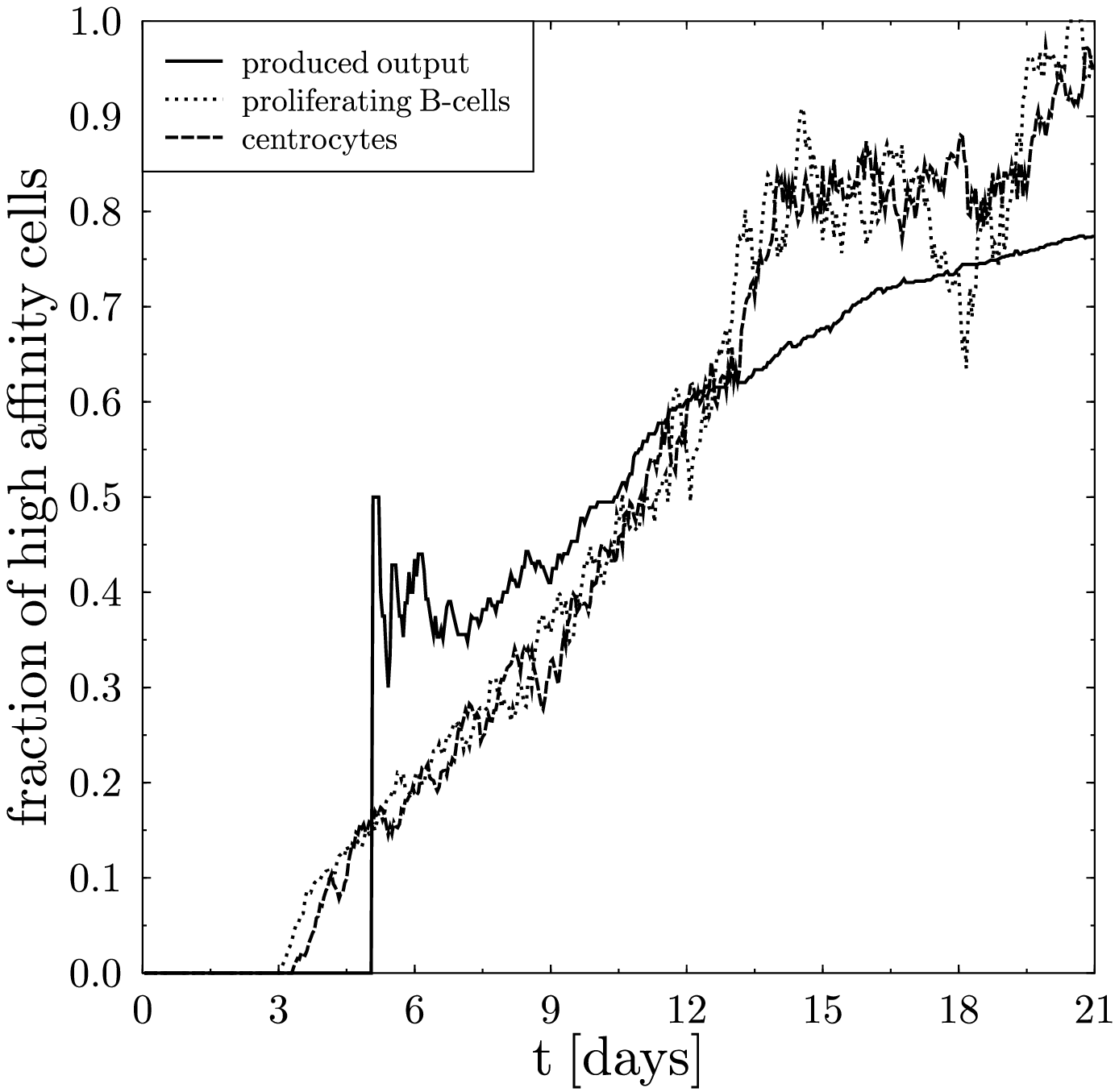}
\end{center}
\vspace*{-5mm}
\caption[]{\sf Affinity enhancement:\\
The time course of the fraction of high
affinity cells (cells which bind the antigen with a probability of at
least $30\%$) in the GC reaction is shown for centroblasts,
centrocytes, and for the sum of all output cells produced
until time $t$ of the GC reaction.}
\label{affinity}
\end{figure}

The average affinity of B-cells is enhanced in four phases during
the whole GC reaction. This is best illustrated by
a typical example with a dark zone present until day $8.3$.
In Fig.~\ref{affinity}
%
the time course of the fraction of high affinity 
centroblasts and centrocytes is shown \cite{mey02}.
{\it High affinity}
denotes those B-cells which bind the antigen with a probability
of more than $30\%$. As can be seen, such a cell does not exist 
at the beginning of the GC reaction. They develop after the
start of somatic hypermutations after $3$ days. Still one
observes a short delay because some mutations have to occur
before the first cells appear that have an above threshold
affinity to the antigen. Then the relative
number of high affinity cells steadily grows. In accordance
with experiment, good cells already dominate around day
$10$ of the GC reaction \cite{jac93}.
One observes an intermediate steep increase, that 
approximately starts
when the dark zone is depleted. This correlation
does not seem to be obvious in view of this single example. 
However, it
has been observed in all simulations, especially
considering different dark zone durations.
In the late phase of the GC
reaction the curve reaches a plateau on a high level,
and the large majority of B-cells are high affinity cells.

In the same figure the fraction of high affinity
output cells is shown. This curve does not show the
value at each moment of the GC reaction but the sum of all
output cells that has been produced until time $t$. 
This value is a measure for the total quality of the
produced output cells.
After day $5$ the quality of the output cells 
is steadily increased during the GC reaction.
The part of high affinity output cells reaches
$77\%$ in the present example. The average over all
simulation with dark zones that vanish between
day $8.5$ and $10.5$ of the reaction is $75.2\%\pm 4.2\%$
(the error is one standard deviation).
In view of the fact that at the beginning of the GC reaction
no high affinity B-cell existed at all, this affinity
enhancement is remarkable.

The main task of the present article is 
to analyze a possible correlation of the duration
of the dark zone and the achieved affinity maturation,
i.e.~the total output quality at the end of the
reaction.
A statistical analysis has to be based on comparable GC simulation.
By changing the production rate of the centroblast differentiation
signal molecule not only the duration of the dark zone
is varied. Also the total duration of the reaction is changed.
As stated at the beginning of this section, the centroblast
differentiation rate is adjusted correspondingly so that
the final numbers of B-cells $N$ after $21$ days of 
the reaction are of the same range.
This is especially important as the affinity maturation
process depends on the value of $N$ \cite{mey02}.
The average number of B-cells for all $226$ simulations
is $\overline{N}=49\,\pm\,23$. 
The error denotes one standard deviation. 
Only those simulations are taken into account that
generate a final number of B-cells $N$ 
within one standard deviation of this average value.

\begin{figure}[ht]
\begin{center}
\includegraphics[height=9cm]{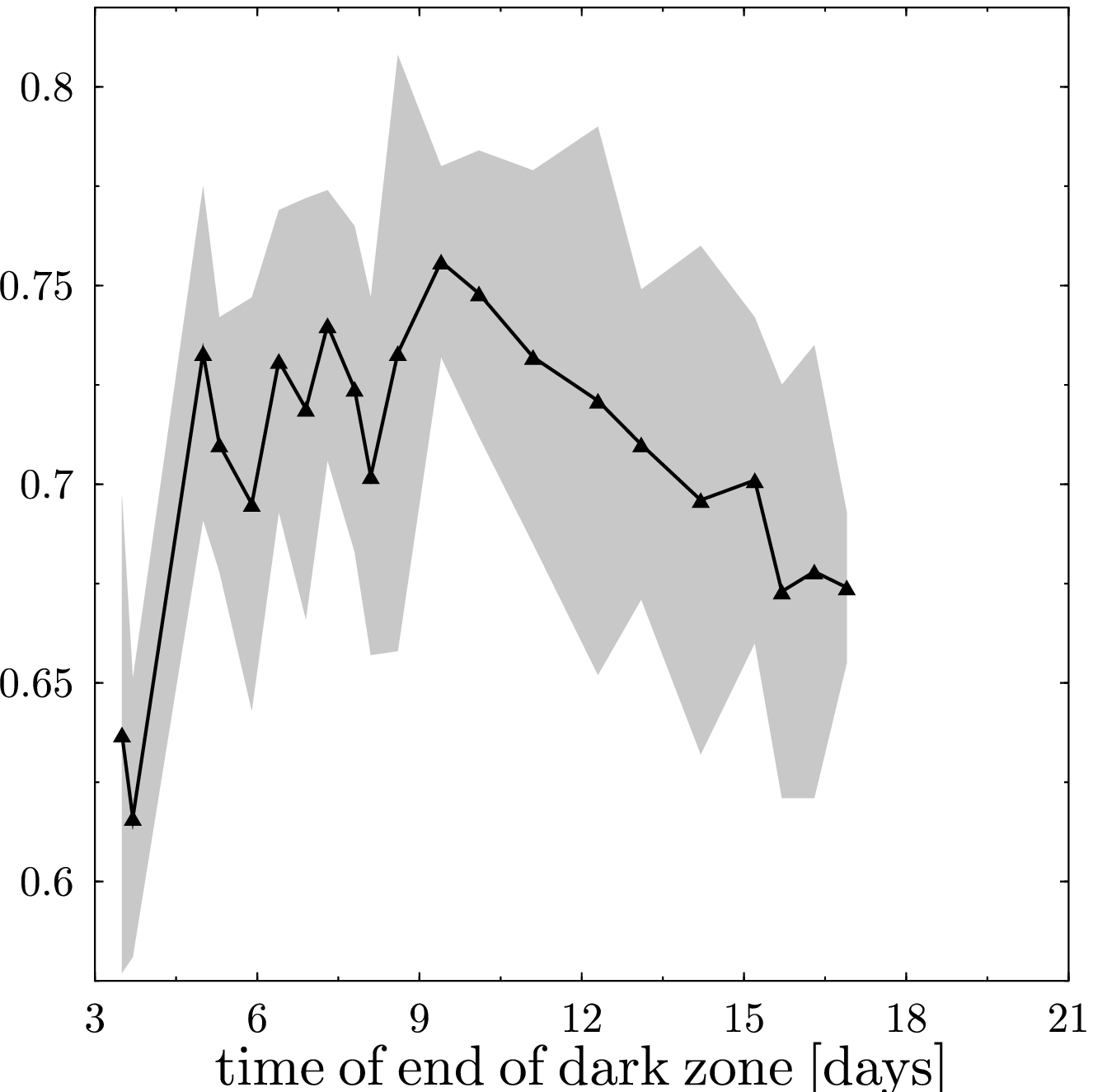}
\end{center}
\caption[]{\sf The importance of the dark zone duration:\\
The dependence of the fraction of high affinity
output cells on the duration of the dark zone. 
The grey area denotes one standard deviation of the average 
values (full line).}
\label{quality}
\end{figure}

The output quality is plotted against the duration of the
dark zone (see Fig.~\ref{quality}). 
%
The quality of the produced
output cells becomes optimized for dark zones that
vanish between day $7$ and $11$ of the GC reaction.
For shorter and longer dark zones, the resulting
quality of the output cells is reduced.
The same holds true for the total number of output cells
(data not shown).
It is worth pointing out, that intermediate dark zones
vanishing around day 9 optimize affinity maturation
on the level of quantity and quality of output cells.


\section{Discussion}
\label{discuss}

We used a previously developed model to elucidate a
possible correlation between the duration of the
dark zone and the efficiency of the affinity maturation process.
In a first step the simulated GCs were compared
to GCs observed in experiment.
The essential properties of real GC reactions were
correctly reproduced. This includes the
appearance of the dark zone, the time course of 
the total volume, as well as the reached
affinity maturation. 

It has been previously found \cite{mey02} that dark zones
do not appear in simulations that are based on
local cell interactions only.
Therefore, it was necessary to introduce an non-local
cell interaction into the model in order to understand
the intermediate appearance of dark zones as observed
in experiment \cite{cam98}.

A centroblast differentiation
signal molecule has been assumed 
that is secreted by FDCs and consumed 
by the centroblasts.
It has been previously shown that such a signal is a
promising candidate to explain appearance and
depletion of dark zones in GCs \cite{mey02}.
However, the signal molecule has to be understood as a
hypothesis, because it is not known how the
centroblast differentiation process is initiated in
real GC reaction.
There are experimental hints that centroblast, indeed, differentiate
due to an interaction with FDCs or T-helper 
cells \cite{han95b,dub99a}.

Most parameters in the model are strongly constrained by
experimental data and consequently the variation of
parameters is restricted. However, the
production of the hypothesized differentiation signal
does not underly such a restriction.
It turned out that the duration of dark zones
strongly depends on the production rate of the differentiation
signal and that other model parameters are less important. 
This situation opens the possibility to vary the
signal production rate and in this way to test
the affinity maturation process of GCs with dark zones
of different duration. Note, that a corresponding test is not
possible in experiments.

A statistical analysis of simulations with different
durations of dark zones leads to the conclusion that
the quality
of the output cells averaged over all
produced output cells during the whole GC reaction
is optimized for intermediately appearing dark zones
that vanish between day $7$ and $11$ of the GC reaction.
Note that also the
quantity of produced output cells is optimized for
these dark zone durations. 
In addition, the time course of the total GC volume
is in agreement with observed time courses for the
same dark zone durations \cite{mey02}.

This result suggests a relation
of two, at first sight, very different categories:
The GC morphology and affinity maturation of B-cells.
The morphology of the GC is basically 
determined by a non-local interaction with other cell-types.
However, the function of the specific GC morphology
is not restricted to a spatial arrangement of interacting
cells. One may suspect that specific cell arrangements
are advantageous for the success of the GC reaction.
The duration of the dark zone does not only determine 
the time course of the cell distribution in GCs 
but also to be a critical parameter for the quality and amount
of output of the GC reaction. This leads to the question
of how the spatial distribution of centroblasts and
centrocytes influences the affinity maturation process.

On one hand the
existence of the dark zone is necessary 
to produce a large pool of different B-cell types,
i.e.~a high diversity of encoded antibody-types.
This diversification is optimally realized with
a fast proliferation of centroblasts without major
interaction to antigens. 
The mutations are likely to occur randomly \cite{rad98}
and it is supposed that starting from low (but non-vanishing)
affinity seeder cells better B-cells are found during
this process, eventhough most clones will be of even less
quality from a probabilistic point of view. 
The diversification process continues
in the dark zone when the selection process has already
started. Note that recycled B-cells that return into
the dark zone don't have been observed in the simulations \cite{mey02}.
Therefore, the B-cells that proliferate in the dark zone
are not recycled B-cells but directly stem from the
original seeder cells.
 
After about $8$ days some high affinity cells have been
found and identified in the selection process
that takes place in the light zone. These roughly
optimized B-cells are mostly recycled cells \cite{vin00} and restart
to proliferate in the light zone. A further enhancement of affinity
to the antigen is based on these preselected B-cells.
How does the ongoing process of diversification
in the dark zone develop? The cells produced in this
random process cannot overcome a certain B-cell quality.
Therefore, a re-proliferation of already selected B-cells
is more promising for the further development of the
GC reaction. This process resembles a directioned
selection process that replaces the random process
in the dark zone. Ongoing proliferation and mutation
in the dark zone does not only become senseless (as the
quality of the cells in the dark zone are below
the average B-cell quality in the light zone). They
would also hinder the optimization process in the light
zone because B-cells of relatively low affinity would
take part in the selection process. Consequently, the selection
of high affinity cells would be inhibited by
a repetition of a first selection process
with B-cells stemming from the dark zone. 
An early depletion of the dark zone eliminates the
low affinty B-cells from the GC and in this way
allows a fine-tuning of preselected B-cells.
We conclude that a dark zone is needed in an early stage
of the GC reaction to give 
centroblasts the possibility to proliferate and mutate
independently of the FDC network. At later stages
a fine-tuning of already preselected B-cells in the direct 
neighborhood of the antigen presenting FDCs becomes
more important and a persisting dark zone inhibits
affinity maturation.

One should be aware
that this interpretation only provides a possible explanation
of the correlation between morphology and affinity
maturation that has been found in the framework
of a spatial model for GCs. 
The basic concepts used here are in accordance with a widely
accepted picture of GCs. Nevertheless, the used
method has to be critically reviewed. The simulation are
based on a regular lattice. We can widely exclude an effect
of lattice symmetries on the results by comparing results
from different lattice architectures \cite{bey02}.
We have presented a statistical evaluation of two-dimensional
simulations. Indeed, the general behaviour of the simulations
does not change in three dimensions \cite{mey03}.

A more relevant problem is the representation of cells of different
size on equal lattice nodes. We have shown that such a description
leads to wrong results if considering chemotaxis as the driving
mechanism for the development of the dark zone \cite{bey02}. 
In such a scenario typical cell velocities differ in two orders
of magnitude and a more reliable description of cell volumes
becomes unavoidable. We have therfore tested if the results
are altered using this more sophisticated cell volume concept.
For cells as slow as in the present simulations the more simple
volume concept turns out to be completely sufficient (data not
shown).

In regular lattice simulations the number of neighboring
lattice nodes is fixed. In a one-cell-one-node description
this determines the number of possible interaction partners
of a cell. The effect of this restriction is not easily
estimated. The simulation should be compared to a lattice
free description. To this end we have implemented the first
three dimensional dynamical Delaunay triangulation code with variable
number of vertices \cite{sch03}. The cells are identified with
the vertices (in a continuous space) 
and the Delaunay triangulation is used in order
to calculate the neighbors of each cell. The triangulation
has to be maintained after the movement of cells (vertices).
In order to allow for proliferation and apoptosis the creation
and deletion of vertices has to be included. The application
of this method to GC reaction will provide another test for
the stability of the results presented here.


\subsection*{Acknowledgments}
I thank Tilo Beyer, Andreas Deutsch and Gernot Schaller
for intense discussions and valuable comments.



\begin{thebibliography}{99}

\bibitem{kro87}
\normalsize
{\sc Kroese, F.G.\,/\,Wubbena, A.S.\,/\,Seijen, H.G.\,/\,Nieuwenhuis, 
P.},
\normalsize
{\rm 1987}:
\normalsize
{\rm Germinal centers develop oligoclonally},
\normalsize
{\it Eur. J. Immunol.\/}
\normalsize
{\bf 17},
\normalsize
{\rm 1069-1072}.
\normalsize

\bibitem{jac93}
\normalsize
{\sc Jacob, J.\,/\,Przylepa, J.\,/\,Miller, C.\,/\,Kelsoe, G.},
\normalsize
{\rm 1993}:
\normalsize
{\rm In situ studies of the primary response to (4-hydroxy-3-nitrophenyl)acetyl. 
III. The kinetics of V region mutation and selection 
in germinal center B cells},
\normalsize
{\it J. Exp. Med.\/}
\normalsize
{\bf 178},
\normalsize
{\rm 1293-1307}.
\normalsize

\bibitem{han97}
\normalsize
{\sc Han, S.\,/\,Zheng, B.\,/\,Takahashi, Y.\,/\,Kelsoe, 
G.},
\normalsize
{\rm 1997}:
\normalsize
{\rm Distinctive characteristics of germinal center B cells},
\normalsize
{\it Immunology\/}
\normalsize
{\bf 9},
\normalsize
{\rm 255-260}.
\normalsize

\bibitem{liu89}
\normalsize
{\sc Liu, Y.J.\,/\,Joshua, D.E.\,/\,Williams, G.T.\,/\,Smith, C.A.\,/\,Gordon, 
J.\,/ MacLennan, I.C.},
\normalsize
{\rm 1989}:
\normalsize
{\rm Mechanism of antigen-driven selection in germinal centres},
\normalsize
{\it Nature\/}
\normalsize
{\bf 342},
\normalsize
{\rm 929-931}.
\normalsize

\bibitem{eij01}
\normalsize
{\sc van Eijk, M.\,/\,Medema, J.P.\,/\,de Groot, C.},
\normalsize
{\rm 2001}:
\normalsize
{\rm Cellular Fas-Associated Death Domain-Like IL-1-Converting 
Enzyme-Inhibitory Protein Protects Germinal Center 
B Cells from Apoptosis Durin Germinal Center Reactions},
\normalsize
{\it J. Immunol.\/}
\normalsize
{\bf 166},
\normalsize
{\rm 6473-6476}.
\normalsize

\bibitem{lin97}
\normalsize
{\sc Lindhout, E.\,/\,Koopman, G.\,/\,Pals, S.T.\,/\,de 
Groot, C.},
\normalsize
{\rm 1997}:
\normalsize
{\rm Triple check for antigen specificity of B cells during 
germinal centre reactions},
\normalsize
{\it Immunol. Today\/}
\normalsize
{\bf 18},
\normalsize
{\rm 573-576}.
\normalsize

\bibitem{liu91}
\normalsize
{\sc Liu, Y.J.\,/\,Zhang, J.\,/\,Lane, P.J.\,/\,Chan, E.Y.\,/\,MacLennan, 
I.C.M.},
\normalsize
{\rm 1991}:
\normalsize
{\rm Sites of specific B cell activation in primary and 
secondary responses to T cell-dependent and T cell-independent 
antigens},
\normalsize
{\it Eur. J. Immunol.\/}
\normalsize
{\bf 21},
\normalsize
{\rm 2951-2962}.
\normalsize

\bibitem{cam98}
\normalsize
{\sc Camacho, S.A.\,/\,Koscovilbois, M.H.\,/\,Berek, C.},
\normalsize
{\rm 1998}:
\normalsize
{\rm The Dynamic Structure of the Germinal Center},
\normalsize
{\it Immunol. Today\/}
\normalsize
{\bf 19},
\normalsize
{\rm 511-514}.
\normalsize

\bibitem{mey02}
\normalsize
{\sc Meyer-Hermann, M.},
\normalsize
{\rm 2002}:
\normalsize
{\rm A Mathematical Model for the Germinal Center Morphology 
and Affinity Maturation},
\normalsize
{\it J. Theor. Biol.\/}
\normalsize
{\bf 216},
\normalsize
{\rm 273-300}.
\normalsize

\bibitem{mey01}
\normalsize
{\sc Meyer-Hermann, M.\,/\,Deutsch, A.\,/\,Or-Guil, M.},
\normalsize
{\rm 2001}:
\normalsize
{\rm Recycling Probability and Dynamical Properties of Germinal 
Center Reactions},
\normalsize
{\it J. Theor. Biol.\/}
\normalsize
{\bf 210},
\normalsize
{\rm 265-285}.
\normalsize

\bibitem{liu94}
\normalsize
{\sc Liu, Y.J.\,/\,Barthelemy, C.\,/\,de Bouteiller, O.\,/\,Banchereau, 
J.},
\normalsize
{\rm 1994}:
\normalsize
{\rm The differences in survival and phenotype between centroblasts 
and centrocytes},
\normalsize
{\it Adv. Exp. Med. Biol.\/}
\normalsize
{\bf 355},
\normalsize
{\rm 213-218}.
\normalsize

\bibitem{bey02}
\normalsize
{\sc Beyer, T.\,/\,Meyer-Hermann, M.\,/\,Soff, G.},
\normalsize
{\rm 2002}:
\normalsize
{\rm A possible role of chemotaxis in germinal center formation},
\normalsize
{\it Int. Immunol.\/}
{\bf 14}, {\rm 1369-1381}.
\normalsize

\bibitem{han95b}
\normalsize
{\sc Han, S.H.\,/\,Hathcock, K.\,/\,Zheng, B.\,/\,Kepler, T.B.\,/\,Hodes, 
R.\,/ Kelsoe, G.},
\normalsize
{\rm 1995}:
\normalsize
{\rm Cellular Interaction in Germinal Centers: Roles of 
CD40-Ligand and B7-1 and B7-2 in Established Germinal 
Centers},
\normalsize
{\it J. Immunol.\/}
\normalsize
{\bf 155},
\normalsize
{\rm 556-567}.
\normalsize

\bibitem{per79}
\normalsize
{\sc Perelson, A.S.\,/\,Oster, G.F.},
\normalsize
{\rm 1979}:
\normalsize
{\rm Theoretical Studies of Clonal Selection: Minimal Antibody 
Repertoire Size and Reliability of Self-Non-self Discrimination},
\normalsize
{\it J. Theor. Biol.\/}
\normalsize
{\bf 81},
\normalsize
{\rm 645-670}.
\normalsize

\bibitem{han64}
\normalsize
{\sc Hanna, M.G.},
\normalsize
{\rm 1964}:
\normalsize
{\rm An autoradiographic study of the germinal center in 
spleen white pulp during early intervals of the immune 
response},
\normalsize
{\it Lab. Invest.\/}
\normalsize
{\bf 13},
\normalsize
{\rm 95-104}.
\normalsize

\bibitem{nos91}
\normalsize
{\sc Nossal, G.},
\normalsize
{\rm 1991}:
\normalsize
{\rm The molecular and cellular basis of affinity maturation 
in the antibody response},
\normalsize
{\it Cell\/}
\normalsize
{\bf 68},
\normalsize
{\rm 1-2}.
\normalsize

\bibitem{cho00}
\normalsize
{\sc Choe, J.\,/\,Li, L.\,/\,Zhang, X.\,/\,Gregory, 
C.D.\,/\,Choi, Y.S.},
\normalsize
{\rm 2000}:
\normalsize
{\rm Distinct Role of Follicular Dendritic Cells and T Cells 
in the Proliferation, Differentiation, and Apoptosis 
of a Centroblast Cell Line, L3055},
\normalsize
{\it J. Immunol.\/}
\normalsize
{\bf 164},
\normalsize
{\rm 56-63}.
\normalsize

\bibitem{eij99}
\normalsize
{\sc van Eijk, M.\,/\,de Groot, C.},
\normalsize
{\rm 1999}:
\normalsize
{\rm Germinal Center B-Cell Apoptosis Requires Both Caspase 
and Cathepsin Activity},
\normalsize
{\it J. Immunol.\/}
\normalsize
{\bf 163},
\normalsize
{\rm 2478-2482}.
\normalsize

\bibitem{wed97}
\normalsize
{\sc Wedemayer, G.J.\,/\,Patten, P.A.\,/\,Wang, L.H.\,/\,Schultz, 
P.G.\,/ Stevens, R.C.},
\normalsize
{\rm 1997}:
\normalsize
{\rm Structural insights into the evolution of an antibody 
combining site},
\normalsize
{\it Science\/}
\normalsize
{\bf 276},
\normalsize
{\rm 1665-1669}.
\normalsize

\bibitem{hol92}
\normalsize
{\sc Hollowood, K.\,/\,Macartney, J.},
\normalsize
{\rm 1992}:
\normalsize
{\rm Cell kinetics of the germinal center reaction --- a 
stathmokinetic study},
\normalsize
{\it Eur. J. Immunol.\/}
\normalsize
{\bf 22},
\normalsize
{\rm 261-266}.
\normalsize

\bibitem{dub99a}
\normalsize
{\sc Dubois, B.\,/\,Barth\'el\'emy, C.\,/\,Durand, I.\,/\,Liu, Y.-J.\,/\,Caux, 
C.\,/ Bri\`ere, F.},
\normalsize
{\rm 1999}:
\normalsize
{\rm Toward a Role of Dendritic Cells in the Germinal Center 
Reaction -- Triggering of B-Cell Proliferation and Isotype 
Switching},
\normalsize
{\it J. Immunol.\/}
\normalsize
{\bf 162},
\normalsize
{\rm 3428-3436}.
\normalsize

\bibitem{rad98}
\normalsize
{\sc Radmacher, M.D.\,/\,Kelsoe, G.\,/\,Kepler, T.B.},
\normalsize
{\rm 1998}:
\normalsize
{\rm Predicted and Inferred Waiting-Times for Key Mutations 
in the Germinal Center Reaction -- Evidence for Stochasticity 
in Selection},
\normalsize
{\it Immunol. Cell Biol.\/}
\normalsize
{\bf 76},
\normalsize
{\rm 373-381}.
\normalsize

\bibitem{vin00}
\normalsize
{\sc de Vinuesa, C.G.\,/\,Cook, M.C.\,/\,Ball, J.\,/\,Drew, M.\,/\,Sunners, 
Y.\,/ Cascalho, M.\,/\,Wabl, M.\,/\,Klaus, G.G.B.\,/\,MacLennan, C.M.},
\normalsize
{\rm 2000}:
\normalsize
{\rm Germinal centers without T cells},
\normalsize
{\it J. Exp. Med.\/}
\normalsize
{\bf 191},
\normalsize
{\rm 485-493}.
\normalsize

\bibitem{mey03}
\normalsize
{\sc Meyer-Hermann, M.\,/\,Beyer, T.},
\normalsize
{\rm 2003}:
\normalsize
{\rm Conclusions from two model concepts on germinal center 
dynamics and morphology},
\normalsize
{\it to appear in Develop. Immunol.\/}
\normalsize

\bibitem{sch03}
\normalsize
{\sc Schaller, G.\,/\,Meyer-Hermann, M.},
\normalsize
{\rm 2003}:
\normalsize
{\rm Dynamic Delaunay tetrahedralizations and Voronoi tessellations 
in three dimensions},
\normalsize
{\it preprint at http://arXiv.org/abs/physics/0302018\/}.
\normalsize




\end{thebibliography}
\end{document}